\documentclass[pra,twocolumn,showpacs,preprintnumbers,amsmath,amssymb]{revtex4}

\usepackage{graphicx}
\usepackage{dcolumn}
\usepackage{bm}
\usepackage{psfrag}
\usepackage{epsfig}
\usepackage{amsmath}
\usepackage{amssymb}
\usepackage{color}
\usepackage[FIGTOPCAP,raggedright,nooneline]{subfigure}

\newcommand{\ignore}[1]{}

\begin{document}

\title{The Bose-Hubbard model with localized particle losses}

\author{Kosmas V. Kepesidis}
\email{kosmas.kepesidis@ph.tum.de}
\author{Michael J. Hartmann}
\affiliation{Department Physik, Technische Universit{\"a}t M{\"u}nchen, James-Franck-Str., 85748 Garching, Germany}

\date{\today}

\begin{abstract}
We consider the Bose-Hubbard model with particle losses at one lattice site. For the non-interacting case, we find that half of the bosons of an initially homogeneous particle distribution, are not affected by dissipation that only acts on one lattice site in the center of the lattice. A physical interpretation of this result is that the surviving particles interfere destructively when they tunnel to the location of the dissipative defect and therefore never reach it. Furthermore we find for a one-dimensional model that a fraction of the particles can propagate across the dissipative defect even if the rate of tunneling between adjacent lattice sites is much slower than the loss rate at the defect.
We analyze the robustness of our findings with respect to small interactions and small deviations from the symmetric setting.
A possible experimental realization of our setup is provided by ultracold bosonic atoms in an optical lattice, where an electron beam on a single lattice site ionizes atoms that are then extracted by an electrostatic field.
\end{abstract}

\maketitle

%
%
\section{Introduction}\label{intro}

Ultracold atoms in optical lattices have in recent years emerged as a very successful quantum simulator for many-particle and solid state physics. The ability to tune parameters of the simulated Hamiltonians provides a unique tool to explore quantum phase transitions in this system \cite{ZwergerRMP}. Among the most studied models is the Bose-Hubbard Hamiltonian which describes interacting bosons in a lattice potential that can tunnel between neighboring lattice sites. In its ground state, it can show two quantum phases,
a superfluid regime for weak interactions and a Mott-insulator phase for strong interactions and commensurate filling \cite{Zwerger}.   
  
Electron beams can provide means for locally probing ultracold quantum gasses since they can be focused onto spot sizes that are much smaller than the optical wavelengths of the trapping fields. High-resolution scanning electron microscopy thus constitutes a method that combines single atom sensitivity and high spatial resolution \cite{electronbeam2,electronbeam4}.  For this technique, a  focused electron beam scans an optically trapped atomic gas. The atoms are ionized by the electron impact and the produced ions are moved off the dipole trap by an electrostatic field. With this method, single-site addressability of ultracold atomic gases loaded in optical lattices has been demonstrated in one and two dimensional structures \cite{electronbeam4,electronbeam3}. In addition, second and higher order correlation functions of a trapped gas of bosons have been extracted \cite{Ott1loss}. 

Besides providing a tool for local measurements, a focused electron beam can also be used as a method to introduce controlled localized dissipation. Along these lines, the formation of solitons in a continuous system with a localized dissipative perturbation has been discussed employing a mean field approach \cite{Ott2loss} and corrections beyond the mean field level have also been considered \cite{theory3}. Moreover, a three-site Bose-Hubbard model with dissipation in the central lattice site has been analyzed \cite{russians} and
a numerical study of the dynamics during a finite time range that employed DMRG methods has been presented \cite{kollath}.
 
In this article we investigate the effect of a localized dissipative defect in a Bose-Hubbard model with analytical means.
This allows us to consider arbitrarily large lattices as well as the long time limit of the dynamics.
For the non-interacting case, we show that a certain fraction of the population of a uniform particle distribution is not affected by dissipation that is localized in the center of the trap. In other words, we show the existence of a dissipation-free subspace for localized particle losses.

Furthermore we find for a one-dimensional model that a fraction of the particles can propagate across the dissipative defect even if the rate of tunneling between adjacent lattice sites is much slower than the loss rate at the defect. Particles can however only tunnel through the dissipative defect at the expense of other particles being lost via the localized dissipation.
 
 For the description of the system we employ a Born-Markov master equation in the standard Lindblad form \cite{WMbook,Lambrobook,Lindmath,Gorinimath}. This equation represents the dynamics of the system under the influence of a mechanism that introduces one-by-one particle losses at one lattice site.
 
 We start by investigating a non-interacting system with the defect located exactly at the center of the chain. In this case, we show the existence of a dissipation-free subspace which consists of the half of the normal modes of the Hamiltonian. Those are the modes that correspond to wave-functions which are anti-symmetric with respect to the trap center. The bosons of those modes interfere destructively at the location of the defect and therefore never reach the lossy site. Since their wavefunction is finite everywhere else, these bosons can tunnel between the right and the left sides of the system without appearing at the defect. 
 
 After this observation, we investigate cases that might present dynamical behavior similar to that of the non-interacting and symmetric one. Most important candidates are cases that correspond to small deviations from the non-interacting and symmetric case. One way to introduce such a deviation is to turn on small interactions while keeping the symmetric structure of the system. Interactions destroy the phase coherence along the lattice since they can transform protected to unprotected modes.
Also in a non-interacting system with a slightly non-symmetric structure, we find that the destructive interference of particles that tunnel into the defect and therefore the dissipation free subspace are perturbed. For those two types of small deviations from the non-interacting and symmetric case, namely small interactions and slightly non-symmetric structure, we show that there is a slow down of the losses for some fraction of the total population.

The remainder of the paper is organized as follows. In section II we introduce our model and discuss general properties of particles in trapping potentials that are symmetric around a dissipative defect. In section III, we present quantitative results for non-interacting bosons on a lattice with a dissipative defect in the central lattice site. Section IV then discusses small deviations from the interaction-free and perfectly symmetric case and we finish the presentation with a summary and conclusions in section V.

\section{Localized losses in symmetrical potentials}\label{sec:gen-discuss}

Various quantum systems can feature states that are protected against certain types of dissipation. Those states - often referred to as dark states - are well known in atomic physics and quantum optics \cite{WMbook,Lambrobook}. In many-body systems such states can appear as a property of a symmetry of a given Hamiltonian and be part of its eigenstates. One of the symmetries that can generate such a dissipation-free subspace is the symmetry under reflection at the origin of the system. In other words, when the potential is symmetric around a localized dissipative defect.

\subsection*{Dissipation-free subspace of a single particle in a symmetrical potential}

We consider a single particle in a symmetric potential. Such a potential is described by an even function $U$ with $ U(x)=U(-x) $. The wavefunctions that correspond to eigenstates of such a potential must have a well defined parity, i.e. some are even  ($ \psi_e(x)=\psi_e(-x) $) and others are odd ($ \psi_o(x)=-\psi_o(-x) $). These properties emerge since the Schr\"odinger equation is unchanged when the sign of a coordinate is reversed \cite{Landau}. If one lists the eigenfunctions according to the number of nodes - increasing energy -  then they are alternately even and odd. The ground state of the system must be an even function since it contains no nodes. Odd functions always vanish at the origin, since
\begin{equation}
\psi_o(0)=-\psi_o(0)=0.
\end{equation}
This means that whatever happens at this particular location in space does not affect the odd eigenfunctions. A physical interpretation of this result is that the probability amplitudes for particles in odd eigenstates destructively interfere at the origin.

Now, let us assume that we locate a mechanism that introduces particles losses at the origin. Then, if the particle is in a odd eigenfunction, it will never be caught by the loss mechanism. This is to say, the odd eigenfunctions constitute a dissipation-free subspace.

\subsection*{Extension to a lattice setup}

\begin{figure}
\includegraphics[width=8cm]{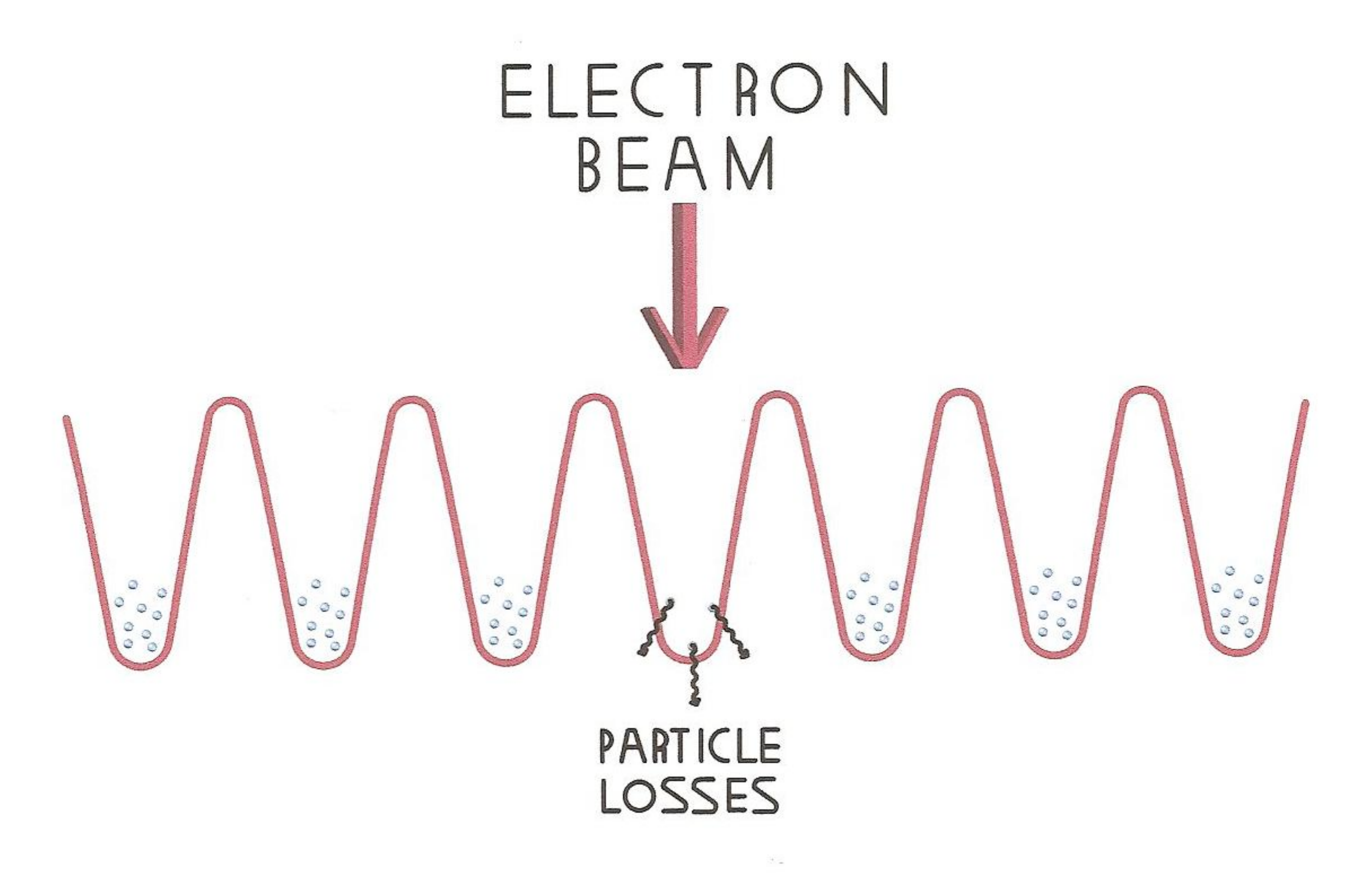}
\caption{\label{symmetricpic} (Color online) Illustration of the proposed setup for the realization of the dissipation-free subspace. Introducing localized particle losses at the central site - assuming odd total number of lattice sites - leaves the odd modes - described by anti-symmetric functions - unaffected. The loss mechanism consists of an electron beam  focused onto the central site. The atom-electron collisions leave the atoms ionized and by the use of an electrostatic field they can be extracted from the lattice. This method provides single-site sensitivity for the control over the position of the defect and high precision control of the associated damping rate.}
\end{figure}

The effect of the dissipation-free subspace may seem (almost) unphysical in a continuous description. This is because we silently assumed that the loss mechanism has zero spatial width. However, this wouldn't be problematic in a lattice setup. If we assume that the relevant physics takes place within the lowest band of a tight binding Hamiltonian, i.e. the energy of the system is low enough, then the lattice sites can be treated as points in space since their extension is much smaller than the wavelength of the mode-functions. Therefore, the only requirement for the loss mechanism is to be restricted to one lattice site, specifically the central one - we assume odd number of sites. Therefore the mechanism that generates the losses needs to be capable of addressing single lattice sites, see Fig.~(\ref{symmetricpic}). Such mechanisms have recently been realized for ultracold atoms in optical lattices \cite{ZwergerRMP,Ott1loss,Ott2loss,Degenfeld12}, which motivates the study we present here.
%

\subsection*{Description of the losses}

We assume that particle losses occur one-by-one and can be described by a super-operator within the Born-Markov approximation \cite{WMbook,Lambrobook} which reads,
\begin{equation}\label{losses}
\hat{\mathcal{L}}_n \left[ \hat{\rho}_S(t) \right] = \frac{\gamma}{2} \left( 2\hat{a}_n \hat{\rho}_S(t) \hat{a}^{\dagger}_n - \hat{a}^{\dagger}_n \hat{a}_n \hat{\rho}_S(t) - \hat{\rho}_S(t)\hat{a}^{\dagger}_n\hat{a}_n \right),
\end{equation}
where $ \hat{\rho}_S(t) $ is the (reduced) density matrix of the system and $ \hat{a}^{\dagger}_n $ and $ \hat{a}_n $ are the creation and annihilation operators of one atom at site $ n $. 
For experimental realizations of local particle losses induced by a focused electron beam, our assumptions for the decay term are well justified.
Here the atoms are ionized by the electron impact and then extracted from the lattice with an electric field. Therefore subsequent particle decays will be independent of each other which justifies the Markov assumption, c.f. \cite{Ott2loss}.

\subsection*{Addition of a confining potential}
In an experimental setup with cold atoms in an optical lattice, the atoms are not only subject to the lattice potential but are also trapped by an additional confining potential that is usually well approximated by a harmonic trap. 
In the central region of the lattice, the value of the confining potential typically only varies very little from one lattice site to another and we therefore approximate it by a constant. In current experiments it is however not guaranteed that the bottom of the confining potential exactly overlaps with the site of the lattice where losses are induced. To estimate the effect of this feature on our findings, we consider deviations from a lattice that is symmetric with respect to the dissipative site in section \ref{notideal}.

\section{Non-interacting bosons in a lattice with particle losses at the central site} \label{ideal}
We start our quantitative study by considering a non-interacting gas of bosonic particles on a lattice with particle losses at the central site. 
Assuming that the tight binding approximation holds, the model under consideration is the Bose-Hubbard Hamiltonian,
\begin{equation}\label{B-H}
\hat{H}=  -J\sum_{\langle i,j \rangle} \left( \hat{a}_{i}^{\dagger}\hat{a}_{j} + h.c. \right) + \frac{U}{2}\sum_{i} \hat{a}_{i}^{\dagger}\hat{a}_i \left( \hat{a}_{i}^{\dagger}\hat{a}_i -1 \right),
\end{equation}
where $ J $ is the tunneling rate of the bosons between neighboring sites and $ U $ is the strength of the (pairwise) on-site particle interaction. The operators $ \hat{a}_i $ and $ \hat{a}^{\dagger}_i $ are the bosonic annihilation and creation operators, respectively. They satisfy the commutation relations for bosons, $ [\hat{a}_i,\hat{a}^{\dagger}_j]=\delta_{ij} $ and $ [\hat{a}_i,\hat{a}_j]=[\hat{a}_i^{\dagger},\hat{a}_j^{\dagger}]=0 $. Since we are interested in the dynamics generated by the Hamiltonian~(\ref{B-H}) for a fixed number of particles, we omit a term related to a chemical potential. Two quantum phases of this model are well known, namely the superfluid and the Mott insulator \cite{Zwerger}. In general the phase of the sample depends on the magnitude of the ratio $ J/U $. An experimental realization of such a model is provided by ultracold bosonic atoms in an optical lattice \cite{ZwergerRMP}. 

Feshbach resonances provide a unique and efficient way to control interactions. Based on this method, an ideal system of non-interacting atoms can be approximately implemented. In this case Eq.~(\ref{B-H}) is approximately written 
\begin{equation}\label{superfluid}
\hat{H}=-J\sum_{\langle i,j \rangle} \left( \hat{a}_{i}^{\dagger}\hat{a}_{i} + h.c. \right).
\end{equation}
This is the hopping Hamiltonian, and because of the absence of interactions the dynamics of each particle is unaffected by presence of the remaining particles.

\subsection{The dynamics in one dimension}

 For simplicity, we start with the dynamics in one dimension. However, as we shall see, the generalization to higher dimensions is straight forward. The Hamiltonian~(\ref{superfluid}) in one dimension is written in the form
\begin{equation}\label{superfluid1D}
\hat{H}=-J\sum_{n=1}^{N-1} \left( \hat{a}_{n}^{\dagger}\hat{a}_{n+1} + h.c. \right),
\end{equation}
where $ N $ is the total number of lattice sites.
We assume open boundary conditions, which imply that the wavefunction vanishes at the edges of the system. The ladder operators for the normal modes, $ \hat{b}_k $, are in this case given by the following transform
\begin{equation}\label{obc}\begin{split}
\hat{a}_n=\sqrt{\frac{2}{N+1}}\sum_{k}\sin \left(nk\right) \hat{b}_k, & \\
\hat{b}_k=\sqrt{\frac{2}{N+1}}\sum_{n=1}^{N}\sin \left(nk\right) \hat{a}_n,
\end{split}\end{equation}
where the quasi-momentum $ k $ is defined as $ k=\frac{\pi l}{N+1} $, with $l=1,2,.....,N $.
It can be easily shown that the operators $ \hat{b}_k $ and $ \hat{b}_k^{\dagger} $ satisfy the commutation relations for bosons. By substituting Eq.~(\ref{obc}) into Eq.~(\ref{superfluid}), we can write the Hamiltonian in the diagonal form
\begin{equation}\label{H-O}
\hat{H}=-2J\sum_{k}\cos \left(k \right) \hat{b}_{k}^{\dagger}\hat{b}_k.
\end{equation}
This Hamiltonian corresponds to a set of $ N $ harmonic oscillators with frequencies given by the dispersion relation $ \omega(k)=-2J\cos \left(k \right) $.

We note that even though the momentum modes (\ref{obc}) only diagonalize the Hamiltonian (\ref{superfluid1D}) in the absence of a confining potential, there is a different set of normal modes which diagonalizes the model with harmonic confining potential as there are no interactions. Following our discussion in section \ref{sec:gen-discuss} half of those normal modes will nonetheless be protected against the localized dissipation.

\subsection*{The Master Equation}
The Born-Markov master equation that describes the evolution of the system under the influence of the dissipation at one lattice site is of the form,
\begin{equation}\label{M-E}
 \frac{d}{dt}\hat{\rho}_S(t)=  -i\left[ \hat{H},\hat{\rho}_S(t) \right] + \hat{\mathcal{L}}_m \left[ \hat{\rho}_S(t) \right]
\end{equation}
where we have set $ \hbar = 1$. The dissipator in the above master equation describes particle losses at site $ m $, as defined in Eq~(\ref{losses}). In a first approximation, we shall assume that this site corresponds to the central site of the lattice. This assumption of a symmetric structure is motivated by the fact that the lattice site on which the electron beam is  focused can in experiments be selected with great accuracy \cite{electronbeam4,electronbeam3}. 

In terms of the mode operators given by Eq.~(\ref{obc}), the dissipator of the master equation reads 
\begin{equation}\label{Momentum}\begin{split}
 \mathcal{\hat{L}}_m \left[ \hat{\rho}_S(t) \right]  = &  \frac{\gamma}{N+1} \sum_{k,p} \sin \left( m k \right) \sin \left( m p \right) \{ 2\hat{b}_{k}\hat{\rho}_S(t)\hat{b}_p^{\dagger}  \\
 &  - \hat{b}_k^{\dagger}\hat{b}_p\hat{\rho}_S(t) - \hat{\rho}_S(t)\hat{b}_k^{\dagger}\hat{b}_p \}
\end{split} \end{equation}
Under the assumption that the total number of sites is odd, the central site is numbered as $ m=(N+1)/2 $ and by using the definition of the quasi-momentum, the dissipator in Eq.~(\ref{Momentum}) is written
\begin{equation}\label{Momentum2}\begin{split}
\mathcal{\hat{L}}_m \left[ \hat{\rho}_S(t) \right] = & \frac{\gamma}{N+1} \sum_{l,j=1}^{N} \sin \left( \frac{\pi l}{2} \right) \sin \left( \frac{\pi j}{2} \right) \{ 2\hat{b}_{l}\hat{\rho}_S(t)\hat{b}_j^{\dagger}  \\
 &  - \hat{b}_l^{\dagger}\hat{b}_j\hat{\rho}_S(t) - \hat{\rho}_S(t)\hat{b}_l^{\dagger}\hat{b}_j \}
 \end{split}\end{equation}
We immediately see that only odd values of $ l $ and $ j $ contribute to the summation. That means that the dissipation affects only the modes that correspond to odd quasi-momentum labeling number. Therefore, the modes with even $ l $, constitute a dissipation-free subspace of the dissipator as given in Eq.~(\ref{losses}).

At first site, this result seems to be in contradiction with the discussion in the previous section. There, we concluded that the dissipation-free subspace consists of the odd eigenfunctions. However, this contradiction is a misconception since the odd eigenfunctions correspond to even quasi-momentum labeling number and vice versa.  

 The master equation, given by Eq.~(\ref{M-E}) can now be written in the following form,
 \begin{equation}\label{M-E2}\begin{split}
 \frac{d}{dt}\hat{\rho}_S(t)= & -i\left[ \hat{H},\hat{\rho}_S(t) \right]  \\
 & + \frac{\gamma}{N+1} \sum_{l,j=1}^{N} \sin \left( \frac{\pi l}{2} \right) \sin \left( \frac{\pi j}{2} \right) \{ 2\hat{b}_{l}\hat{\rho}_S(t)\hat{b}_j^{\dagger}  \\
 &  - \hat{b}_l^{\dagger}\hat{b}_j\hat{\rho}_S(t) - \hat{\rho}_S(t)\hat{b}_l^{\dagger}\hat{b}_j \}.
  \end{split} \end{equation} 
Using the above equation, one can derive equations governing the evolution of the mean values of observables. For the mean value of the number of bosons in odd modes (with even quasi-momentum labeling number) $ \langle N(t)\rangle_{Odd} = \sum_{l^{\prime}} \langle \hat{b}_{2l^{\prime}}^{\dagger} \hat{b}_{2l^{\prime}} \rangle_{(t)} = \sum_{l^{\prime}} Tr \{ \hat{b}_{2l^{\prime}}^{\dagger} \hat{b}_{2l^{\prime}} \hat{\rho}_S(t) \} $, where $ l^{\prime}=1,2,....,N/2 $, one gets,
 \begin{equation}\label{even}
 \frac{d}{dt} \langle N(t)\rangle_{Odd}=0.
  \end{equation}
 The obvious solution of this equation, $ \langle N(t)\rangle_{Odd} = \langle N(0)\rangle_{Odd}$ 
 confirms that modes with an even quasi-momentum labeling number constitute a dissipation-free subspace.
 
  On the other hand, for the mean value of the number of bosons in even modes (with odd quasi-momentum labeling number) $ \langle N(t)\rangle_{Even} = \sum_{l^{\prime}} \langle \hat{b}_{2l^{\prime}-1}^{\dagger} \hat{b}_{2l^{\prime}-1} \rangle_{(t)} = \sum_{l^{\prime}} Tr \{ \hat{b}_{2l^{\prime}-1}^{\dagger} \hat{b}_{2l^{\prime}-1} \hat{\rho}_S(t) \} $, where $ l^{\prime}=1,2,....,(N+1)/2 $, we have,
\begin{equation}\label{odd}
 \frac{d}{dt} \langle N(t)\rangle_{Even} = -\frac{2 \gamma}{N+1} \sum_{l,j=1}^{(N+1)/2} \langle \hat{b}^{\dagger}_{2j-1} \hat{b}_{2l-1} \rangle_{(t)}
\end{equation}
%
\begin{figure}[tbp]
\includegraphics[width=8cm]{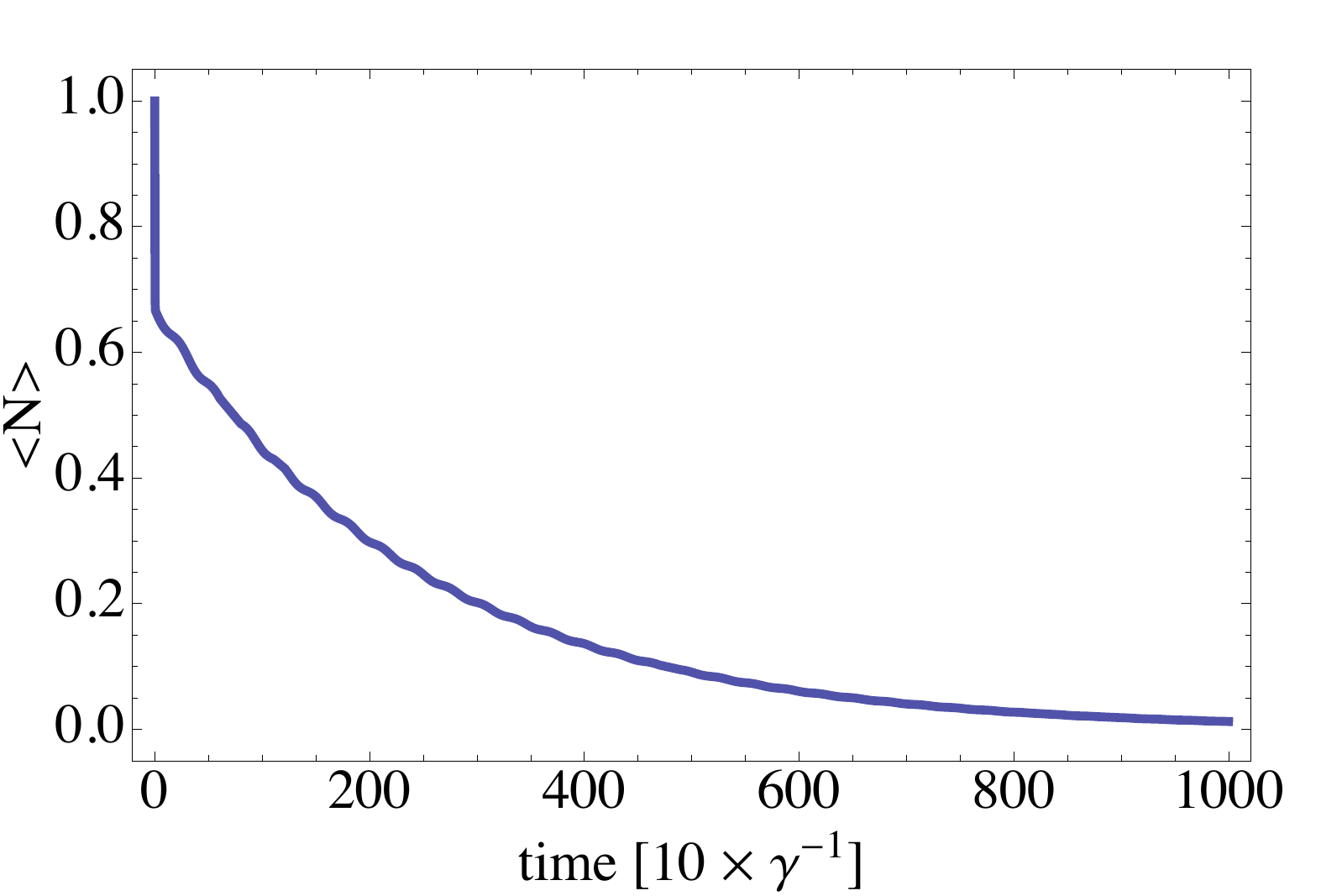}
\caption{\label{evenpic} (Color online) The expectation value of the population of the even modes as a function of time. All the bosons that populate these modes eventually decay. The plot reveals also the rapid loss of the particles initially populating the defect. A large proportion of the particles is annihilated almost instantly in the case where $ \gamma\gg J $. This suggests that, in this case, the lossy site can be adiabatically eliminated from the description because of its fast rotating dynamics. Parameters:  $ \gamma=10 $, $ J=0.1 $, $ U=0 $ and $ N=5 $.}
\end{figure}
Hence, particles in these modes will eventually be lost from the system.
The numerical solution of the above equation is shown in Fig.~(\ref{evenpic}). One sees that all bosons that populate the even modes eventually decay.

\subsection*{Particle propagation across the defect}

\begin{figure}
\includegraphics[width=8cm]{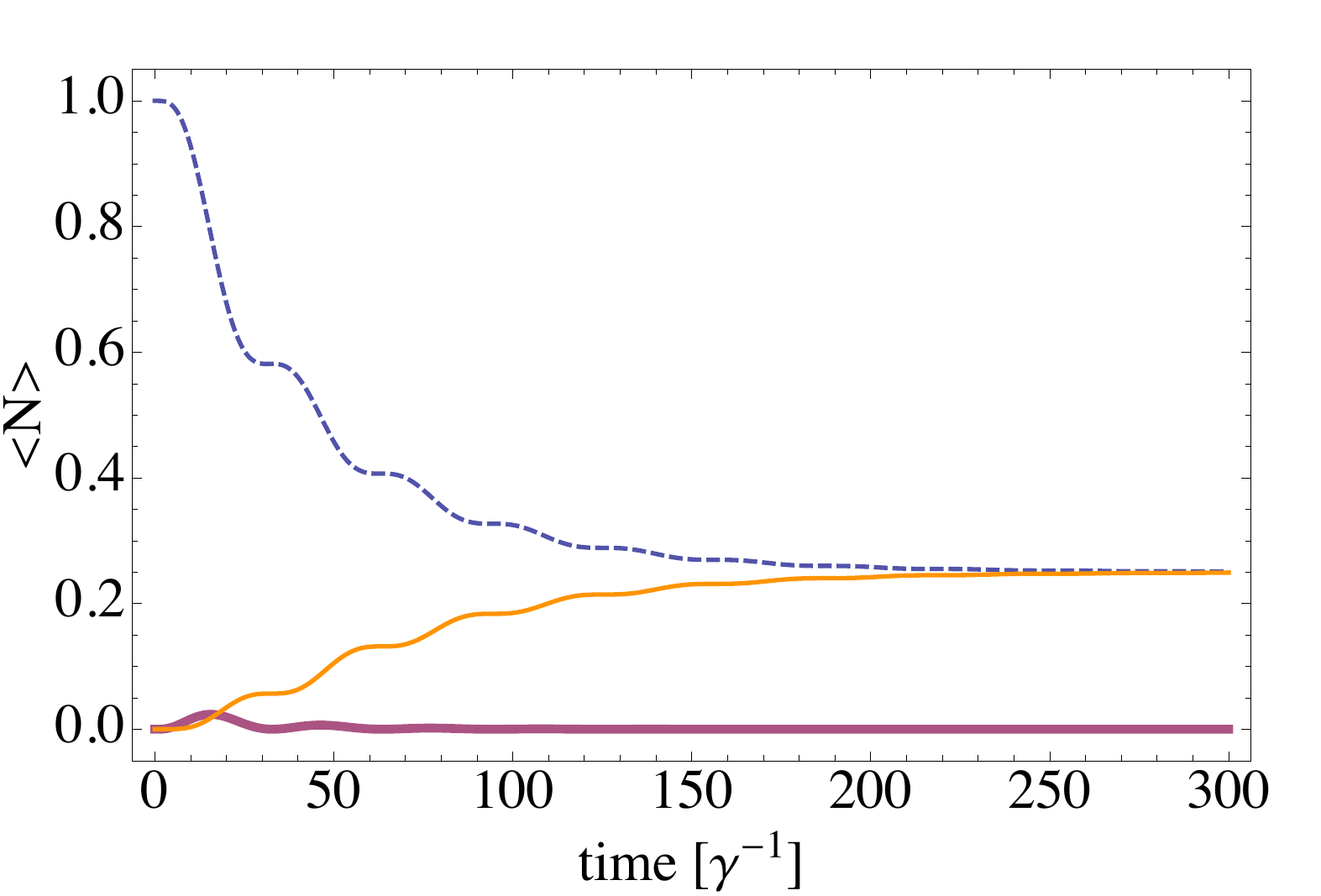}
\caption{\label{propagation} (Color online) The propagation across the defect of a particle that is initially localized at the first lattice site from the left. Dashed line: The expectation value of the total population in lattice sites to the left of the defect. Thin line: The expectation value of the total population in lattice sites to the right of the defect. Thick line: The expectation value of the population in the defect. Half of the initial number of particles survives the damping and ends up in a superposition of being located on the left and the right part of the lattice. Parameters:  $ \gamma=1 $, $ J=0.1 $, $ U=0 $ and $ N=5 $.}
\end{figure}

We now consider the case where initially one boson is located at a specific lattice site at one side of the dissipative defect. 
The state of this localized particle is a coherent superposition of all the eigenmodes of the system \cite{HP08a}. Therefore, the
fraction of the particle that occupies odd modes will be gradually lost due to the dissipation at the defect whereas the remaining part will be protected against the localized losses as it occupies even modes.
Since the even modes however show equal particle densities at both sides of the defect, a fraction of the initially localized particle is able to propagate through the defect into the other half of the lattice, see Fig.~(\ref{propagation}). In complete analogy, a fraction of a gas of non-interacting bosons that is initially located at one side of the dissipative defect, will propagate through the lossy site into the other half of the lattice at the expense of loosing half its initial number of particles, see Fig.~(\ref{propagation}). It is remarkable that this effect is independent of the rate of particle losses at the defect. Hence even if the particles are lost from the defect a lot faster than they can tunnel into it, still a fraction of the particles ($1/4$ of them in our examples) is able to cross the defect and reach the other half of the lattice.
%

\subsection{Extension to higher dimensions}

Since in our case the dissipation-free subspace is a consequence of the symmetry of our system, one expects to find an analogous behavior in higher dimensions.
To this end, we consider a two dimensional square lattice of size $ N \times N $ and suppose that the electron beam focuses exactly on the center of the square lattice. In this case, the Hamiltonian given by Eq.~(\ref{superfluid}) takes on the form,
\begin{equation}\label{2D}
\hat{H}=-J \sum_{i,j=1}^{N-1} \left( \hat{a}_{i,j}^{\dagger} \hat{a}_{i+1,j} + \hat{a}_{i,j}^{\dagger}\hat{a}_{i,j+1} + h.c. \right).
\end{equation}
By assuming again open boundary conditions, the ladder operators are transformed as,
\begin{equation}\label{fourier2D}\begin{split}
& \hat{a}_{i,j}=\frac{2}{N+1}\sum_{k,p} \sin \left( ik \right) \sin \left( jp \right) \hat{b}_{k,p}, \\
& \hat{b}_{k,p}=\frac{2}{N+1}\sum_{i,j=1}^{N} \sin \left( ik \right) \sin \left( jp \right) \hat{a}_{i,j},
\end{split}\end{equation}
where again the quasi-momenta take the values $ k,p=\frac{\pi l}{N+1} $, with $ l=1,2,....,N $. In terms of the mode operators $ \hat{b}_{k,p} $ and $ \hat{b}_{k,p}^{\dagger} $, Hamiltonian~(\ref{2D}) has the following diagonal form,
\begin{equation}\label{superfluid2Ddiag}
\hat{H}=-2J\sum_{k,p} \left[ \cos\left( k \right) + \cos\left( p \right) \right] \hat{b}_{k,p}^{\dagger} \hat{b}_{k,p}.
\end{equation}
The dissipator that describes the particle losses at the central site reads,
\begin{equation}\begin{split}
\hat{\mathcal{L}}_{m,m} \left[ \hat{\rho}_S(t) \right] = & \frac{\gamma}{2} \left( 2\hat{a}_{m,m}\hat{\rho}_S(t)\hat{a}^{\dagger}_{m,m} - \hat{a}^{\dagger}_{m,m}\hat{a}_{m,m}\hat{\rho}_S(t) \right. \\ 
& \left. - \hat{\rho}_S(t)\hat{a}^{\dagger}_{m,m}\hat{a}_{m,m} \right)
\end{split}\end{equation}
where the coordinates of the lossy site are given by $ m=(N+1)/2 $. In terms of the eigenmode operators defined in Eq.~(\ref{fourier2D}), the above dissipator reads,
\begin{equation}\begin{split}
\hat{\mathcal{L}}_{m,m} \left[ \hat{\rho}_S(t) \right] & =  \frac{2\gamma}{(N+1)^2}\\
\times & \sum_{i,j,l,f=1}^{N} \sin\left( \frac{\pi i}{2} \right)\sin\left(\frac{\pi j}{2}\right) \sin\left(\frac{\pi l}{2}\right)\sin\left(\frac{\pi f}{2}\right) \\
\times &\{ 2\hat{b}_{i,j} \hat{\rho}_S(t) \hat{b}^{\dagger}_{l,f} - \hat{b}^{\dagger}_{i,j}\hat{b}_{l,f}\hat{\rho}_S(t) - \hat{\rho}_S(t)\hat{b}^{\dagger}_{i,j}\hat{b}_{l,f} \}.
\end{split}\end{equation}
One immediately sees that the above dissipator is non-zero only when both the entries of the summations are odd (corresponding to even modes). The eigenmodes that correspond to any other combination of quasi-momentum labeling numbers, constitute part of the dissipation-free subspace. This can be demonstrated by computing the evolution of the total number of particles in modes corresponding to every such combination. For the modes where at least one index of quasi-mometum labeling is an even number, one obtains,
\begin{equation}\label{zeros}
 \frac{d}{dt}\langle N(t) \rangle_{O-O}=\frac{d}{dt}\langle N(t) \rangle_{E-O} = 0. 
\end{equation}
On the other hand for the modes where both indices are odd numbers (even modes) one gets, 
\begin{equation}\label{odd-odd}
\frac{d}{dt} \langle N(t) \rangle_{E-E}  =  - \frac{4\gamma}{(N+1)^2}
\sum_{l,j,k,f=1}^{(N+1)/2} \langle \hat{b}^{\dagger}_{2l-1,2j-1} \hat{b}_{2k-1,2f-1} \rangle_{(t)} .
\end{equation}
We therefore conclude that the dissipation-free subspace of a two dimensional non-interacting, symmetric system, constitutes the $3/4$ of all its eigenmodes. 

In a completely analogous way, one finds that in a three-dimensional structure, the protected modes are $ 7/8 $ of all its eigenmodes.

\subsection{The case of large damping rate and the adiabatic elimination of the defect}

Let us return to the one dimensional setup. As illustrated in Fig.~(\ref{evenpic}), in the case where the damping rate $ \gamma $ is much larger than the tunneling $ J $, the loss of the particles that populate the defect, occurs almost instantly compared to the time-scale in which the tunneling takes place. This suggests that one can assume the lossy site to be empty at all times and adiabatically eliminate its degrees of freedom to generate an effective picture for the description of the remainder of the system. We now discuss this approach.

\subsection*{The effective Markovian dissipator}

We consider a system with three sites \cite{russians} with particle losses occurring at the central site. By performing perturbative theory in the interaction picture \cite{WMbook,Lambrobook} and tracing out the degrees of freedom of the lossy site, we derive an expression for the evolution of the reduced density matrix of two outer sites $\hat{\rho}_{-m}$,
\begin{equation}\label{rdm}
\hat{\mathcal{L}}_{eff} \left[\hat{\rho}_{-m}(t)\right] = - \int_{0}^{t} \text{Tr}_{m}\left[ \hat{V}(t), \left[ \hat{V}(t-s),\hat{\rho}(s) \right] \right] ds.
\end{equation}
Since we trace out the degrees of freedom of the defect, we automatically place the lossy site in the status of a quantum environment. Thus, the interaction Hamiltonian $ \hat{V}(t) $ in Eq.~(\ref{rdm}) corresponds to the tunneling between the defect and its neighbor sites,
\begin{equation}
\hat{V}(t)=-J \left( \hat{a}^{\dagger}_{m-1}(t) \hat{a}_m(t) + \hat{a}^{\dagger}_{m+1}(t) \hat{a}_m(t) + h.c. \right).
\end{equation}
We now focus on the regime $ \gamma \gg J $ where we can assume that the lossy site practically remains empty at all times, as illustrated in  Fig.~(\ref{evenpic}). By imposing the Born approximation the equation (\ref{rdm}) reads,
\begin{equation}\begin{split}\label{rdm2}
\hat{\mathcal{L}}_{eff} \left[\hat{\rho}_{-m}(t) \right]= &  - \int_{0}^{t} \text{Tr}_{m}\left[ \hat{V}(t), \left[ \hat{V}(t-s), \right.\right. \\ 
& \left.\left. |0\rangle \langle0| \otimes \hat{\rho}_{-m}(s) \right] \right] ds,
\end{split}\end{equation}
for this regime. The integral kernel in the above equation contains the two-time correlation functions, $ \langle0| \hat{a}_m(t)\hat{a}^{\dagger}_m(s) |0\rangle $ and $ \langle0| \hat{a}_m(s)\hat{a}^{\dagger}_m(t) |0\rangle $, of the defect. These functions can be calculated using the dissipator given by Eq.~(\ref{losses}) which, in this case, yields the internal dynamics of the defect. By doing so and using the quantum regression theorem \cite{WMbook,Lambrobook}, one finds that the two-time correlations decay exponentially,
\begin{equation}
\langle0| \hat{a}_m(t)\hat{a}^{\dagger}_m(t-s) |0\rangle = \langle0| \hat{a}_m(t-s)\hat{a}^{\dagger}_m(t) |0\rangle = e^{-\frac{\gamma}{2}s}.
\end{equation}
The characteristic time-scale in which the correlations decay is $ \tau=2/\gamma $. In the case where the damping rate is large, the correlations decay rapidly, as illustrated in Fig.~(\ref{evenpic}). Since the correlations are short-lived, the interactions occurs sharply at $ s=t $. Therefore, it is safe to replace $ s $ in $ \hat{\rho}_{-m}(s) $ in Eq.~(\ref{rdm2}), by $ t $. And by extending the upper limit of the integration to infinity, which concludes the Markov approximation, we can perform the integration with respect to $ s $, obtaining,
\begin{equation}\begin{split}\label{generalized}
\hat{\mathcal{L}}_{eff}\left[\hat{\rho}_{-m}(t)\right] & = \frac{2J^2}{\gamma} \left( 2 \tilde{a}_{m} \hat{\rho}_{-m}(t) \tilde{a}_{m}^{\dagger} \right.\\
 & \left. - \tilde{a}_{m}^{\dagger} \tilde{a}_{m} \hat{\rho}_{-m}(t) - \hat{\rho}_{-m}(t)  \tilde{a}_{m}^{\dagger} \tilde{a}_{m} \right).
\end{split}
\end{equation}
where $\tilde{a}_{m} = \hat{a}_{m-1}+\hat{a}_{m+1}$. Hence, the dynamics of the lossy site is no longer part of this effective description and the losses are described as a property of certain type of states, namely symmetric superpositions between the neighbors of the defect. 

\subsection*{The dissipation-free subspace in the effective picture}
Since the damping affects only symmetric superpositions, states that do not overlap with them are immune to particle losses. Therefore, the anti-symmetric states of the from,
\begin{equation}
|\psi\rangle_A = \frac{1}{\sqrt{n!}} \left(\frac{\hat{a}^{\dagger}_{m-1} - \hat{a}^{\dagger}_{m+1}}{\sqrt{2}} \right)^{n} |0\rangle,
\end{equation}
for any integer number $n$ have constant population. It can be analytically shown that those anti-symmetric states identically equal to the odd modes that constitute the dissipation-free subspace.

\subsection*{Relation to the Quantum Zeno effect}
Another implication of Eq.~(\ref{generalized}) comes from the structure of the effective dumping rate $ 2J^2/\gamma $. As the damping rate $ \gamma $ increases, the actual losses decrease. This behavior is a manifestation of the quantum Zeno effect \cite{Zenotheory,Zeno+}. This result has been experimentally discovered for ultra-cold molecules by Syassen et al. \cite{Zenoexp} and is predicted in\cite{russians,kollath}, see also \cite{Witthaut09,Trimborn11}. Although this effect is not directly obvious from the form of the dissipator given by Eq.~(\ref{Momentum2}), the related phenomenon is of course not a consequence of the adiabatic elimination. 

\section{Interactions and asymmetry: Two factors that induce losses to the dissipation-free subspace} \label{notideal}

All the above discussion was limited to the case of a symmetric structure (defect at the center of the lattice) and no interactions. In this section we investigate how deviations from this ideal case can affect the protected modes. First we shall consider the case of a symmetric structure but with small yet finite interactions. Then we shall check the dynamics when the defect is not located at the center of the lattice but rather at a random position along it.  

\subsection{Small interactions}

Phenomena that depend on interference are typically affected by the presence of interactions, since the latter destroy the phase coherence. This happens because the term in the Bose-Hubbard Hamiltonian (\ref{B-H}), which is responsible for the interactions, is non-linear and gives rise
to a non-linear term in the corresponding Schr\"odinger equation. However, in the case of small enough interactions, one expects the system to behave similarly to the non-interacting case, at least for sort times. In order to examine to what extent interactions affect our findings of the previous sections, we compare the strength of the interactions to the damping rate that appears in Eq.~(\ref{odd}) for an 1D lattice.

For a one-dimensional Bose-Hubbard system, the part of the Hamiltonian~(\ref{B-H}) that describes the on-site particle interaction,
in terms of the mode operators defined in Eq.~(\ref{obc}), reads,
\begin{equation}\label{int-modes}
\hat{H}_{int} = \frac{U}{2} \sum_{i,j,l,f} \mathcal{T}(i,j,l,f) \hat{b}^{\dagger}_k\hat{b}_p\hat{b}^{\dagger}_l\hat{b}_f - \frac{U}{2} \sum_{i,j} \mathcal{S}(i,j) \hat{b}^{\dagger}_i\hat{b}_j,
\end{equation}
with
\begin{equation}\begin{split}
 \mathcal{T}(k,p,l,f) = & \frac{4}{(N+1)^2} \sum_{n=1}^{N}  \sin \left( \frac{n\pi i}{N+1} \right)\sin \left( \frac{n\pi j}{N+1} \right) \\
 & \times \sin \left( \frac{n\pi l}{N+1} \right)\sin \left( \frac{n\pi f}{N+1} \right)
\end{split}\end{equation} 
 and 
 \begin{equation} 
 \mathcal{S}(i,j) = \frac{2}{N+1} \sum_{n=1}^{N} \sin \left( \frac{n\pi i}{N+1} \right ) \sin \left( \frac{n\pi j}{N+1} \right ).
 \end{equation} 
One can show that function $ \mathcal{S} $ is identically equal to a delta function, $ \mathcal{S}(i,j)= \delta_{i,j } $. This means that the second term of the right-hand side of Eq.~(\ref{int-modes}) is diagonal and generates no hopping of particles between protected and unprotected modes. On the contrary, the function $ \mathcal{T} $ is not a delta function and therefore the non-linear term of the Hamiltonian~(\ref{int-modes}) is responsible for hopping between the two types of modes.

\begin{figure}[tbp]
\includegraphics[width=8cm]{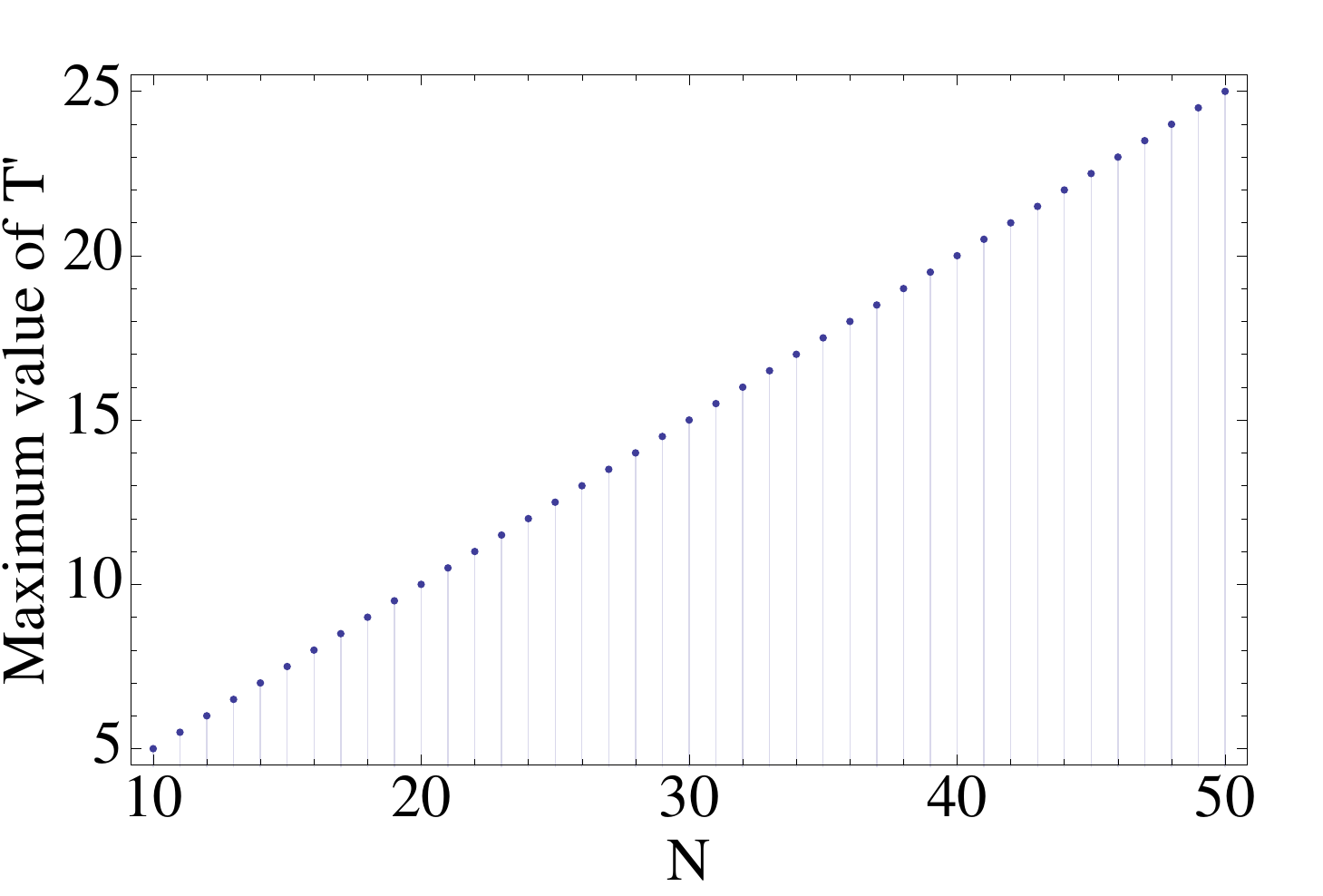}
\caption{\label{max} (Color online) The maximum value of $ T^{\prime} $ as a function of the size of the system - the total number of lattice sites $N$. The plot reveals the linear character of the function. Specifically, $ \text{Max}[T^{\prime}]\sim N/2 $.}
\end{figure}

For small enough interactions, the scattering of particles from the protected into the unprotected modes occurs a lot slower than the subsequent loss of particles from the unprotected modes. Hence, one expects a that particles in the protected modes still survive a lot longer than the other particles. To find a condition for when this scenario occurs, we consider the ratio between the scattering rate $ U \mathcal{T} / 2$ and the effective damping rate $ \Gamma = 2 \gamma/(N+1) $ as given in Eq.~(\ref{odd}). The scattering from protected to unprotected modes needs to be slower than the losses for all modes involved. We therefore calculate when,
\begin{equation}\label{eff-tun}
\frac{U}{2 \Gamma} \, \text{Max} [|\mathcal{T}|] \ll 1 .
\end{equation} 
To this end, we numerically calculate the maximum value of $ T^{\prime}(i,j,l,f) = \frac{(N+1)^2}{4} \mathcal{T}(i,j,l,f) $ as a function of the size of the system, $N$, see Fig.~(\ref{max}), and observe that $ \max[T^{\prime}] \approx N/2 $. 
Therefore, the ratio of the maximum value of the scattering over the effective damping is,
\begin{equation}
\frac{U}{2 \Gamma} \, \text{Max} [|\mathcal{T}|] \approx \frac{UN}{2 \gamma (N+1)} \sim \frac{U}{\gamma}.
\end{equation}
We thus conclude that the slow-down takes place in the regime where $ \gamma \gg U $.
An analogous treatment shows that this condition is also valid in two dimensions. 

\subsection{Deviations from the symmetry}

Up to now, we have assumed that particle losses occur at the center of the chain, see Eq.~(\ref{M-E}). Now, we shall place the losses not at the central site $ m $ but rather to a random position along the lattice. Instead of subscript $ m $, we use $ r $ which stands for "random". 
From the structure of the normal modes in equation (\ref{obc}) one can see that for each location $r$, those modes with vanishing amplitude at $r$ are protected against losses affecting the site $r$ only.

\begin{figure}[tbp]
\begin{center}
\includegraphics[width=9cm]{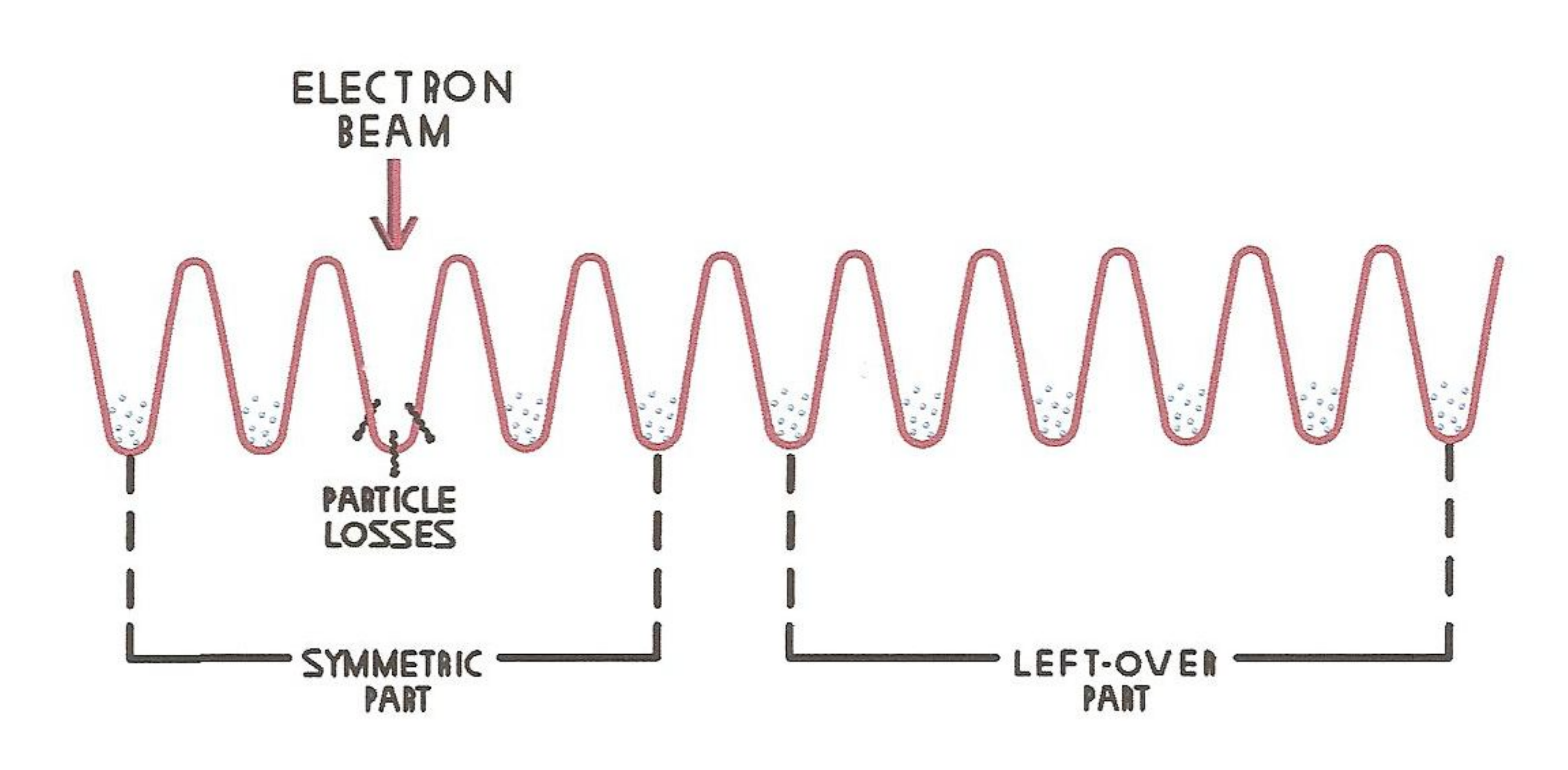}
\caption{\label{asymmetricpic} (Color online) Illustration of an asymmetric setup. The defect is located at a random position along the lattice rather than being at the center. For the description of the dynamics, we split the lattice into two parts, - each part is described by a different set of modes - the symmetric one that consists of the sites around the defect, obtaining in this way the structure of the symmetric setup, and the one consisting of the left-over sites. }
\end{center}
\end{figure}

A general tendency to what degree the dissipation free subspace is degraded by deviations from the symmetry of the setup can be obtained from the following consideration.
We assume, without loss of generality, that the defect is located at a site on the left of the center of the lattice. We can split the lattice into two parts, the symmetric part around the defect and the remaining part that consists of the sites that make the right side of the lattice larger - an illustration is given in Fig.~(\ref{asymmetricpic}).

In this way, we can again retain modes that are symmetry-related. Specifically, for the symmetric part of the lattice, namely for the first site from the left to site number $2r-1$, we define the modes,
\begin{equation}\begin{split}\label{sym-modes}
& \hat{a}_n =\frac{1}{\sqrt{r}} \sum_{l=1}^{2r-1} \sin \left( \frac{\pi n l}{2r} \right) \hat{b}_l \\
& \hat{b}_l =\frac{1}{\sqrt{r}} \sum_{n=1}^{2r-1} \sin \left( \frac{\pi n l}{2r} \right) \hat{a}_n.
\end{split}\end{equation}
Accordingly, for the left-over sites, namely for site $ 2r $ to site $ N $, we define the additional modes,
\begin{equation}\begin{split}\label{left-over}
& \hat{a}_n =\frac{2}{\sqrt{N-2r}} \sum_{l=2r}^{N} \sin \left( \frac{\pi n l}{N-2r} \right) \hat{c}_l \\
& \hat{c}_l =\frac{2}{\sqrt{N-2r}} \sum_{n=2r}^{N} \sin \left( \frac{\pi n l}{N-2r} \right) \hat{a}_n.
\end{split}\end{equation}
Since the lossy site $ r $ belongs to the symmetric part of the lattice the ladder operators $ \hat{a}^{\dagger}_r $ and $ \hat{a}_r $ can be expanded in the mode operators defined in Eq.~(\ref{sym-modes}). Therefore, the dissipator given by Eq.~(\ref{losses}), reads in this case,
\begin{equation}\begin{split}\label{rand-mod}
\hat{\mathcal{L}}_{r} \left[\hat{\rho}_S(t)\right] = & \frac{\gamma}{2r} \sum_{l,j=1}^{2r-1} \sin \left( \frac{\pi l}{2} \right) \sin \left( \frac{\pi j}{2} \right) \{ 2\hat{b}_l \hat{\rho}_S(t) \hat{b}^{\dagger}_j \\
& - \hat{b}^{\dagger}_l \hat{b}_j \hat{\rho}_S(t) - \hat{\rho}_S(t) \hat{b}^{\dagger}_l \hat{b}_j \}.
\end{split}\end{equation} 
Similarly to the ideal case, we easily observe that the above dissipator affects only the modes which correspond to odd quasi-momentum labeling number. That means that the modes with even quasi-momentum labeling number that correspond to the symmetric part and all the modes that correspond to the left-over part are not directly affected by the particle losses.

Since we consider again a non-interacting case, the unitary part of the evolution of the system is governed by the Hamiltonian given by Eq.~(\ref{superfluid1D}). In order to write this Hamiltonian in terms of the mode operators (\ref{sym-modes}) and (\ref{left-over}), we split it into three parts,
\begin{equation}\begin{split}\label{Ham-split}
\hat{H} = & -J \sum_{n=1}^{2r-2} \left( \hat{a}^{\dagger}_{n}\hat{a}_{n+1} + h.c. \right) - J \sum_{n=2r}^{N-1} \left( \hat{a}^{\dagger}_{n}\hat{a}_{n+1} + h.c. \right) \\
& - J \left( \hat{a}^{\dagger}_{2r-1}\hat{a}_{2r} + h.c. \right).
\end{split}\end{equation}   
The first part of the above Hamiltonian describes the symmetric part of the system, the second describes the left-over sites and the last one gives the tunneling between the last site of the symmetric part and the first one of the left-over part. Now, in terms of the modes (\ref{sym-modes}) and (\ref{left-over}), the Hamiltonian reads,
\begin{equation}\begin{split}\label{Ham-split-modes}
\hat{H} = & -2J \sum_{l=1}^{2r-1} \cos \left( \frac{\pi l}{2r} \right) \hat{b}^{\dagger}_{l}\hat{b}_{l} -2J \sum_{j=2r}^{N} \cos \left( \frac{\pi j}{N-2r} \right) \hat{c}^{\dagger}_{j}\hat{c}_{j} \\
& - J \sum_{l=1}^{2r-1} \sum_{j=2r}^{N} \mathcal{R}(j,l)  \{ \hat{b}^{\dagger}_{l}\hat{c}_{j} + h.c. \},
\end{split}\end{equation}
where we have defined,
\begin{equation}
\mathcal{R}(j,l) = \frac{2}{\sqrt{2r(N-2r)}} \sin \left( \frac{\pi (2r-1) l}{2r} \right) \sin \left( \frac{\pi 2r j}{N-2r} \right)
\end{equation}   
The above Hamiltonian includes a tunneling term between the modes of the symmetric part and the left-over ones. This tunneling process can transform protected to unprotected modes and destroys the effect of the dissipation-free subspace. However, if the tunneling between both parts of the lattice occurs slowly compared to the rate of particle losses at the defect, then one expects a slow-down of particle losses for the modes with even quasi-momentum labeling number. In such a case, the system approximately resembles the behavior of the ideal case described in the first section, at least for short times. We therefore analyze this case in a similar way as the case of small interactions. The tunneling of particle from modes with even quasi-momentum labeling number to lossy modes is slow compared to the losses $\Gamma$ affecting the latter provided, 
\begin{equation}\label{slow-down}
\frac{J}{\Gamma} \, \text{Max} [|\mathcal{R}|] \ll 1 .
\end{equation}
Inserting $\Gamma = \gamma/(2 r)$ and the values of $\mathcal{R}$, this ratio reads,
\begin{equation}\label{slow-down1}
\frac{J}{\Gamma} \, \text{Max} [|\mathcal{R}|]  = \frac{J}{\gamma} \sqrt{\frac{2r}{N-2r}}.
\end{equation}
This ratio depends on the relative magnitudes of the parameters $ J $ and $ \gamma $ but also on the size of the asymmetry. 
In the case of small asymmetry, where $  N \simeq 2r  $, the ratio is approximately written,
\begin{equation}\label{slow-down2}
\frac{J}{\Gamma} \, \text{Max} [|\mathcal{R}|]  \approx \frac{J}{\gamma} \sqrt{N}.
\end{equation}
This indicates that there is no slow down of particle losses as long as $ \gamma \gg  J \sqrt{N} $. 
In the opposite case, where the asymmetry is large, i.e. $ N > 2r $, we approximately have,
\begin{equation}\label{assym}
\frac{J}{\Gamma} \, \text{Max} [|\mathcal{R}|]  \approx \frac{J}{\gamma} \sqrt{\frac{2r}{N}}
\end{equation}  
and since the ratio $ 2r /N $ is smaller than one, we have a slow-down if $ \gamma > J $.

\section{conclusion} \label{conclusion}
In this article we have considered the Bose-Hubbard model with particle losses at one lattice site. For the non-interacting case, we found that particles in normal modes with vanishing amplitude at the dissipative defect are not affected by the localized losses. For a one-dimensional model with the lossy site exactly in the center of the chain, half the modes thus form a dissipation free subspace.
This behavior can be attributed to a destructive interference of particles tunneling into the defect.
Furthermore a fraction of the particles can propagate across the dissipative defect even if the rate of tunneling between adjacent lattice sites is much slower than the loss rate at the defect.

To estimate the robustness of the features we predict, we have analyzed the effect of small particle interactions and deviations from a perfectly symmetric setup.

Our findings could be studied experimentally with ultracold bosonic atoms in an optical lattice, where an electron beam on a single lattice site ionizes atoms that are then extracted by an electrostatic field.


\acknowledgements
MJH acknowledges fruitful discussions with Herwig Ott.
This work is part of the Emmy Noether project HA 5593/1-1 and the CRC 631, both funded by the German Research Foundation, DFG.



\begin{thebibliography}{99}
\bibitem{ZwergerRMP}
Immanuel Bloch, Jean Dailibard, Wilhelm Zwerger,
Rev. Mod. Phys. {\bf 80}, 885-964 (2008)
%
\bibitem{Zwerger}
Wilhelm Zwerger,
J. Opt. B: Quantum Semiclass. Opt., {\bf 5}, S9 (2003)
%
\bibitem{electronbeam2}
Dieter Jaksch, Nat. Phys., {\bf 4}, 906 (2008)
%
\bibitem{electronbeam4}
Tatjiana Gericke, Peter W\"urtz, Daniel Reitz, Tim Langen and Herwig Ott, Nat. Phys. {\bf 4}, 949 (2008)
%
\bibitem{electronbeam3}
Peter Wuertz, Tim Langen, Tatjana Gericke, Andreas Koglbauer and Herwig Ott, Phys. Rev. Lett., {\bf 103}, 080404 (2009)
%
\bibitem{Ott1loss}
V. Guarrera, P. W\"urtz, A. Ewerbeck, A. Vogler, G. Barontini and H. Ott,
Phys. Rev. Lett. {\bf 107}, 160403 (2011)
%
\bibitem{Ott2loss}
Valery A. Brazhnyi, Vladimir V. Konotop, Victor M. P\'{e}rez-Garcia and Herwig Ott, 
Phys. Rev. Lett. {\bf 102}, 144101 (2009)
%
\bibitem{theory3}
D. Witthaut, F. Trimborn, H. Hennig, G. Kordas, T. Geisel and S. Wimberger, Phys. Rev. A. {\bf 83}, 063608 (2011)
%
\bibitem{russians}
V.S. Shchesnovich and D.S. Mogilevtsev, Phys. Rev. A {\bf 82}, 
043621 (2010)
%
\bibitem{kollath}
Peter Barmettler and Corinna Kollath, Phys. Rev. A {\bf 84}, 041606(R) (2011)
%
\bibitem{WMbook}
D.F. Walls and G. Milburn, {\it Quantum Optics}, Springer 2010
%
\bibitem{Lambrobook}
P. Lambropoulos and D. Petrosyan, {\it Fundamentals of Quantum Optics and Quantum Information}, Springer 2007
%
\bibitem{Lindmath}
G. Lindblad,
Commun. Math. Phys. {\bf 48}, 119-130 (1976)
%
\bibitem{Gorinimath}
V. Gorini, A. Kassakowskiand E. C. G. Sudarshan,
J. Math. Phys. {\bf 17}, 821-825 (1976)
%
\bibitem{Landau}
L. D. Landau and E. M. Lifshitz, {\it Quantum Mechanics (Non-relativistic Theory)}, Elsevier 2003
%
\bibitem{Degenfeld12} P. Degenfeld-Schonburg, E. del Valle and M.J. Hartmann,
Phys. Rev. A {\bf 85}, 013842 (2012).
%
\bibitem{HP08a} M.J. Hartmann and M.B. Plenio,
Phys. Rev. Lett., {\bf 100}, 070602 (2008).
%
\bibitem{Zenotheory}
B. Misra and E.C.G. Sudarshan, J. Math. Phys. {\bf 18}, 756 (1977)
%
\bibitem{Zeno+}
S. Gammelmark and K. Molmer, Phys. Rev. A {\bf 81}, 012120 (2010)
%
\bibitem{Zenoexp}
N. Syassen, et al., Science {\bf 320}, 1329 (2008)
%
\bibitem{Witthaut09}
D. Witthaut, F. Trimborn, S. Wimberger,
Phys. Rev. A, {\bf 79}, 033621, (2009).

\bibitem{Trimborn11}
F. Trimborn, D. Witthaut, H. Hennig, G. Kordas, T. Geisel, and S. Wimberger,
Eur. Phys. J. D {\bf 63}, 63 (2011).

\end{thebibliography}
\end{document}